# TOWARDS A SYSTEM FOR SEA STATE FORECASTS IN THE BULGARIAN COASTAL ZONE: THE CASE OF THE STORM OF 07-08 FEBRUARY 2012

**Vasko Galabov** 1

**Assoc. Prof. Anna Kortcheva** 1

**Marieta Dimitrova** 1

1 National Institute of Meteorology and Hydrology- Bulgarian Academy of Sciences (NIMH-BAS) - **Bulgaria**

## ABSTRACT

The paper describes the existing operational sea state forecast system of NIMH- BAS for sea state in the Black Sea and our current progress on the implementation of an additional component for the forecasts of wind waves in the Bulgarian coastal zone. Wind Waves and especially the extreme ones, occurring during severe storms are a major hazard for the coastal zone, causing significant damages to the infrastructure, threat for the human lives and also causing significant damages to the protected areas around the coast. The numerical model WAVEWATCH III is in use for wind waves forecasts for the entire Black Sea with horizontal resolution of 1/8 degree ( roughly 14 kilometers), which is sufficient for the open Sea, but not enough for a detailed coastal forecast. For the purposes of the coastal forecasts and early warnings in case of severe storms we decided to implement SWAN (Simulating the Waves Near Shore)- development of TU- DELFT.  In this paper we will describe the brief details about the coastal sea state forecasting system of NIMH- BAS and a case study of the storm of 07-08 February 2012, which is the severest storm in our coastal area for the last decade (and probably longer period).

## INTRODUCTION

Wind waves, together with the strong winds and storm surge events, are major hazards in the coastal areas. The accurate forecast of such events is of a great importance for the national meteorological services in order to be able to issue reliable early warnings to the authorities, general public and users of specialized forecasts such as port and maritime authorities, ships, offshore platforms etc. In Bulgaria the organization, which is responsible to issue such warning on operational basis is the National Institute of Meteorology and Hydrology (NIMH). The wind forecasts, based on numerical modeling, were implemented more than one decade ago- the beginning was in 1997 with the VAGBUL wave model [1]. After that in 2003 WAVEWATCH III [2] was successfully implemented with a spatial resolution of 0.125° (~14km) and since then it provides the wave forecasts to the forecasters of NIMH and external users. WAVEWATCH III (the Black Sea performance of this model, compared with VAG is presented in [3]) is a state of art model for a sea scale forecasts (deep waters), but in order to have an improvement of the coastal forecasts, we realized that in order to improve the coastal warnings in case of storms, it is important to implement also a wave model for the coastal area with higher spatial resolution. For that purpose we implemented the state of art numerical model for coastal waters SWAN (Simulating Waves Near Shore), developed in the technical University of Delft, Netherlands. The model was implemented in pre-operational mode in 2011 and it is fully operational since the autumn of 2011. During the winter of 2011/2012 the coastal wave prediction system was tested for a first stormy season. The most significant storm of the winter- the storm of 06-09 February 2012 caused significant damages around the Bulgarian coast, therefore we decided to focus on this case of severe storm. In order to verify our forecasts for the stormy period at the end of January and beginning of February, we did a comparison of the model output with a satellite altimetry data from JASON and ENVISAT satellites, because there is one major problem- there is no operational in situ data from oceanographic buoys in the Black Sea, which is a major limit not only for us, but to the entire Black Sea meteorological and oceanographic community, nevertheless this method of verification was proven to be reliable enough, even if the amount of data from these satellites is limited due to the small size of the Black Sea.

**THE IMPLEMENTATION OF SWAN MODEL AT NIMH- BAS**

SWAN is a third generation spectral wave model, developed in the technical university of Delft, Netherlands (TU- Delft), [4] with the purpose to be a coastal wave transformation model. Initially, as noted by the authors, the usage of the model was not recommended for larger domains, but with its constant improvement (more specifically the implementation of a high order propagation scheme S&L) the model became applicable even at ocean scales (but not as effective as the widely used Wavewatch III and WAM models). The model is proven to very efficient especially when the spatial resolution goes below 0.1° (below this limit the usage of implicit numerical scheme in SWAN makes it very efficient and economic from the computational point of view and also very flexible). For a complete description of SWAN see [5]. Here we will point out only the main features of the model: SWAN solves the action density balance equation:

The first term on the left side represents the local rate of change of the action density N. Action density N is equal to energy density E, divided by the relative frequency. The second and third term represents the propagation of action in geographic space, with propagation velocities cx and cy .The forth term represents the shifting of the relative frequency due to variations of depth and currents, with propagation velocity cσ in σ-space. The fifth term represents depth-induced and current-induced refraction. The right side of the equation gives the total source term, which for the third generation model is given as: Stot (σ, θ)= Sin (σ, θ) + Snl(σ, θ) + Sds(σ, θ) , where Sin gives the transfer of wind energy to the waves, Snl nonlinear wave-wave interactions and Sds – wave dissipation- here SWAN takes into account the whitecapping (a main source of wave dissipation in deep waters), depth induced wave breaking and bottom friction) (for a complete description of the wave growth and dissipation terms in SWAN see [5]). Regarding the nonlinear wave- wave interactions, SWAN takes into account the quadruplet's mechanism and triads in a very shallow waters.

At NIMH- BAS SWAN was implemented in a spherical grid covering the entire Black Sea with a spatial resolution of 2' (~3.7x2.6 km.), using the implementation of Van der Westhuysen [6] of the saturation based formulation of the whitecapping of Alves and Banner [7] and the wind input of Yan [8]. The reason to use this combination instead of the widely used operationally formulations of Komen and Janssen is our own experience from our operational practice, that it gives better results for the Black Sea and Mediterranean Sea (and probably also all other limited- fetch basins). The model runs twice a day, using the output of ARPEGE/ ALADIN regional atmospheric model, the operational atmospheric model of NIMH- BAS [9], at 06h UTC and 18h UTC, producing a forecasts for the next 72 hours in a form of a graphical output (using GMT software and GrADS software) available in the intranet of the institute (to appear in the web site www.meteo.bg) and as a table output with wave parameters for the next 72 hours for a number of points along the Bulgarian coast. The model don't starts from a flat sea state (without waves) but also not from a saved initial data from a previous run- we initialize the model by doing a stationary run of SWAN at the beginning of the computational cycle and production of a hotstart file for the nonstationary forecast run. We found out that such approach can give better results during the first 6- 12h of the model run, especially if there is a significant difference in the wind fields of two consequent runs of ARPEGE/ ALADIN.

There is also a second domain, nested in the bigger one, which was implemented to produce wave forecasts for the bay of Burgas. The spatial resolution of this domain is 500 meters. An example output from this domain is presented on fig.1.

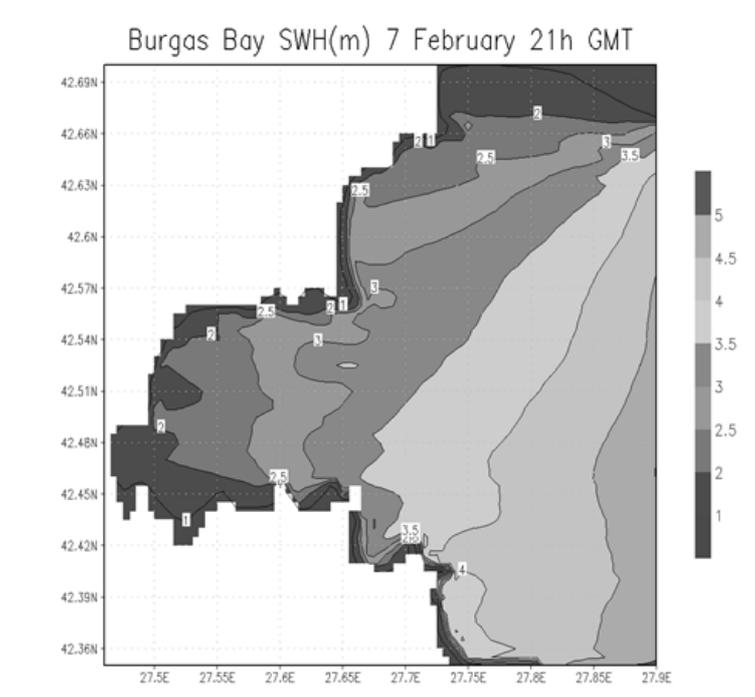

**Fig.1 Example output of the forecast of significant wave height for the bay of Burgas**

The reason to setup also this domain was the fact that the representation of the bay within the model even at 3 kilometers spatial resolution is poor. In the near future we plan also to increase the spatial resolution of the model for the entire Bulgarian coast to one km.

Finally we will point out that there are other authors that have implemented SWAN for the Black Sea and presented important studies, like the studies of E. Rusu [9], [10], but we should point out here, that as far as we know in fact the presented <u>operational</u> implementation of SWAN is the first one for the Black Sea and the first <u>operational</u> system with such high spatial resolution.

**THE STORM OF 07-08 FEBRUARY 2008 AND EVALUATION OF THE PERFORMANCE OF SWAN**

The storm of 07- 08 is of particular importance for a case study, because of the significant damages around the Bulgarian coast. Fig 2 presents the meteorological situation (with huge pressure gradient over the entire Black Sea)

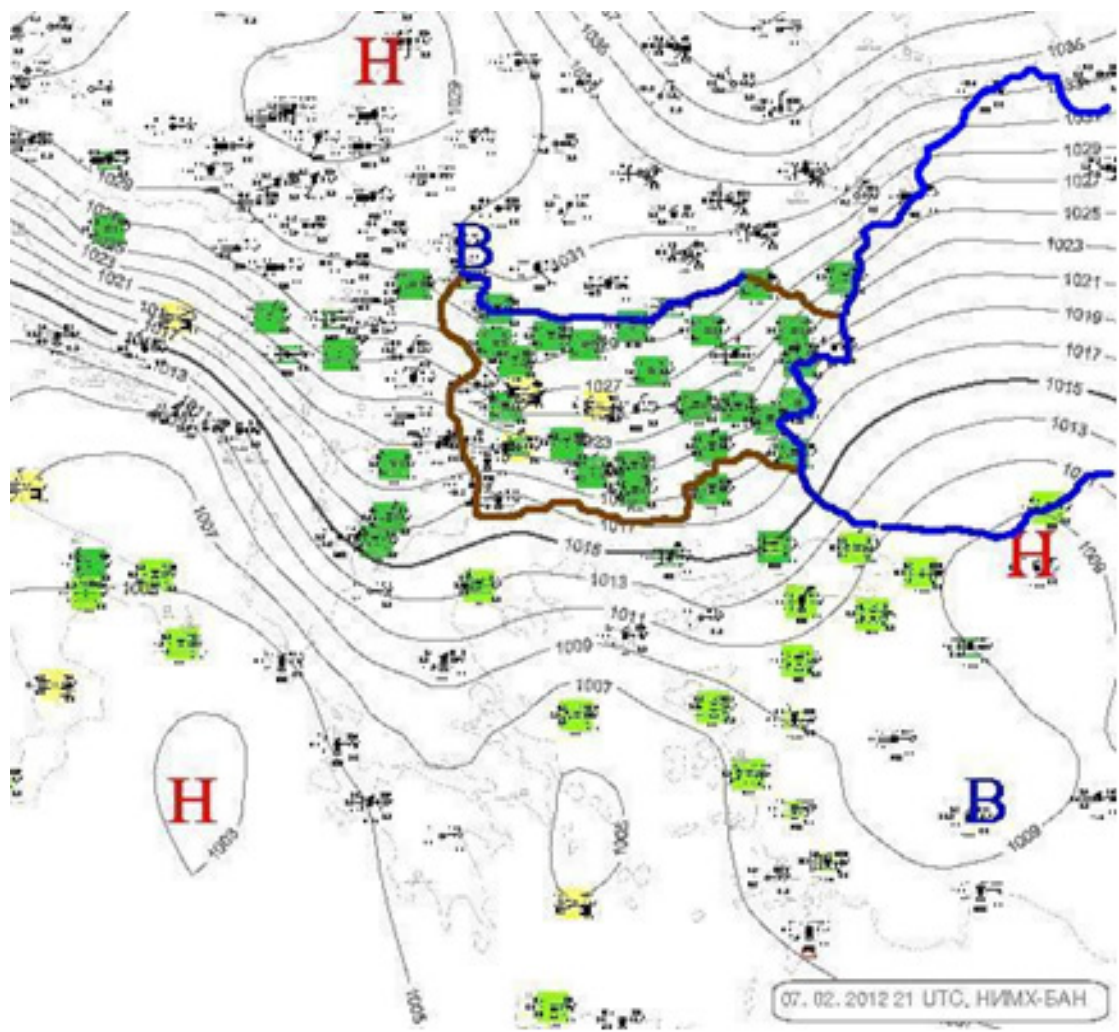

**Fig. 2 Analysis of the synoptic situation at 21h UTC 07.02.2012 (analysis of the forecasters of NIMH- BAS, regional branch Varna, available online at http://varna.meteo.bg/feb-2012.html)**

The wind speed around the coast reached 24- 28m/s with a measurement of 34m/s in one of the stations. In order to evaluate the performance of SWAN, the satellite altimetry data from JASON and ENVISAT satellites was obtained, quality checked (filtered), collocated in the model computational grid and compared with the model results for the period 05-10 February 2012. Fig 3. presents the output of the model during the storm.

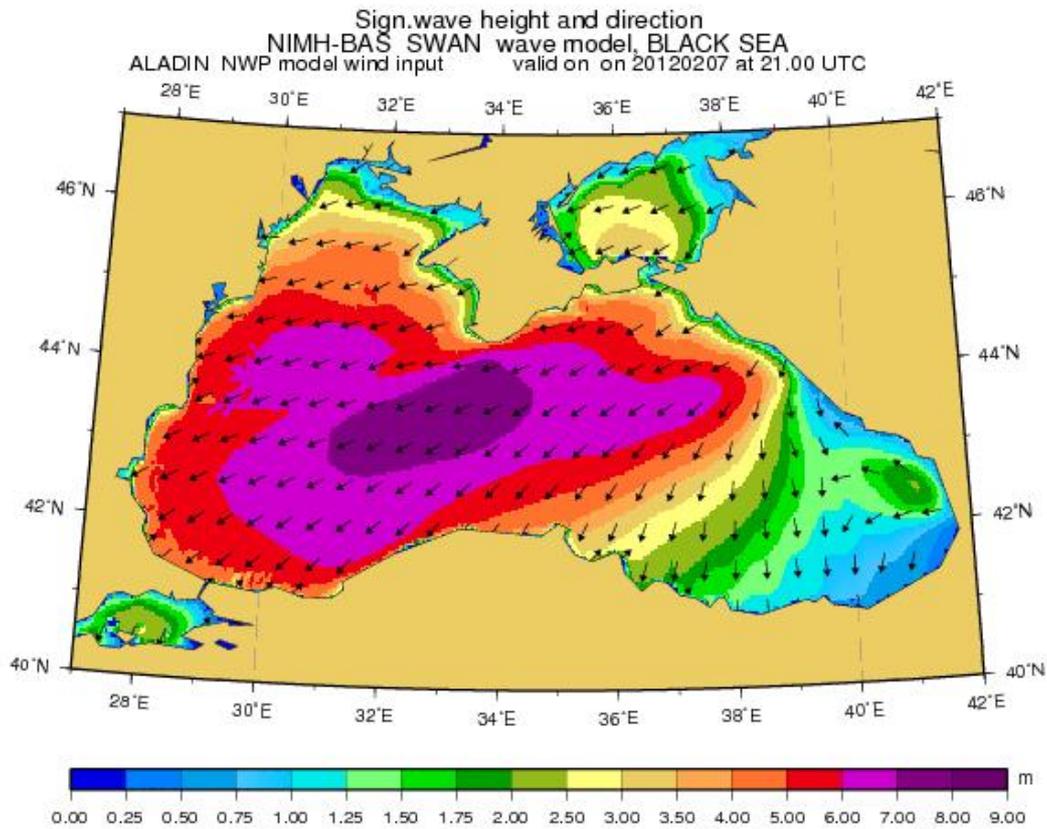

**Fig.3 Output of SWAN (forecast for 07.02.2012 21h UTC)- significant wave height reaching 7-8m and up to 5-6m. In the western shelf waters**

The next figure (4) presents the measured by satellite altimetry significant wave height (output, using Google Earth software). It is easy to observe that the measurement gives a result that corresponds to the model prediction. This way 472 measurement were processed and compared with the model output and table 1 presents the statistics about the model performance.

**Table 1**

| Significant Wave Height [m]- SWAN Black Sea | Bias [m] | Root Mean Square Error (RMSE) [m] | Scatter Index |
|---|---|---|---|
| **Day 1 (+6- +24h)** | -0.17 | 0.45 | 0.10 |
| **Day 2 (+24- +48h)** | -0.36 | 0.68 | 0.13 |
| **Day 3 (+48- +72h)** | -0.27 | 0.65 | 0.14 |
| **Entire period of forecast (+6- +72h)** | -0.28 | 0.61 | 0.13 |

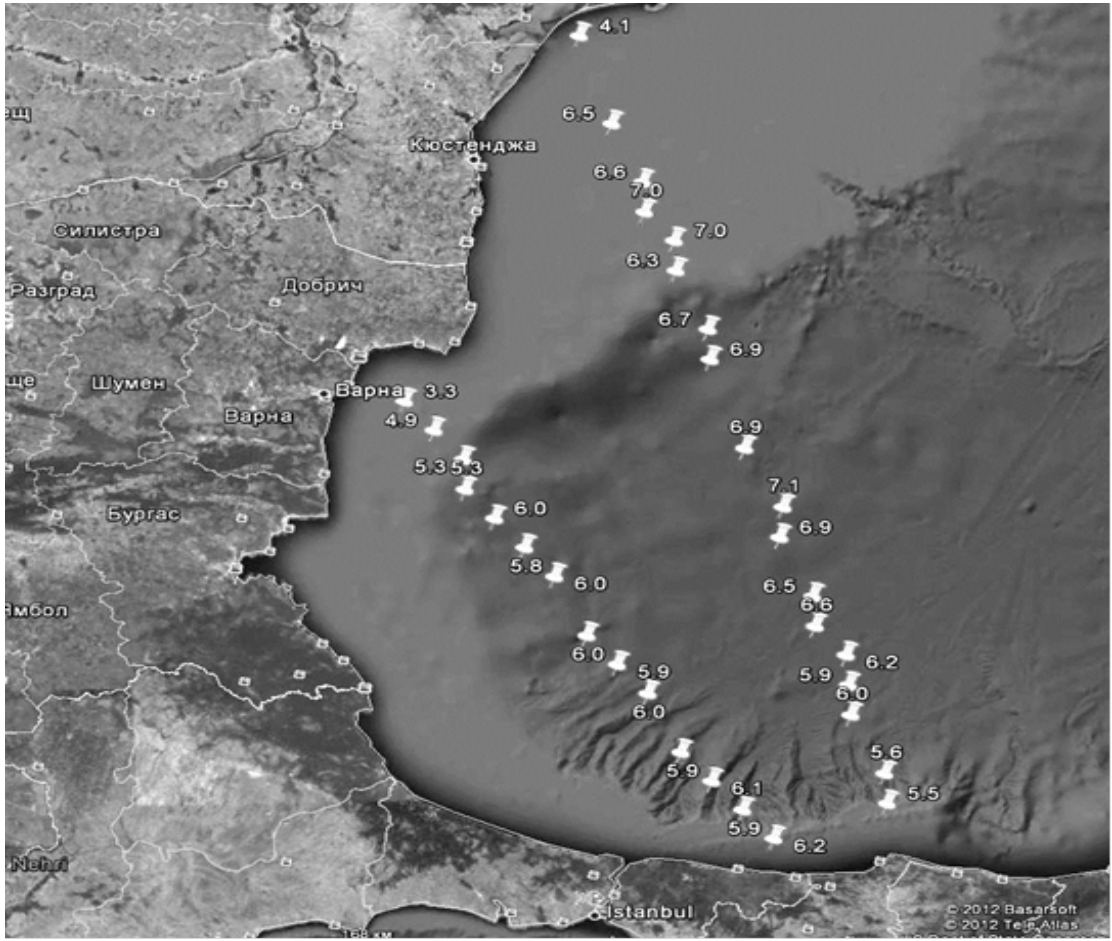

**Fig. 4 Significant wave height (m), measured by the ENVISAT satellite on 07.02.2012 21h UTC- the right track and 08.02.2012 by JASON1 satellite- the left track**

The bias was found to be ~ -0.2 -04m and the root mean square error below 0.5 for the first 24h forecast and below 0.7 for the second and third day, which is a good result, taking into account that we are comparing not for typical, but for extreme rare events. In general the under prediction by the model was found to be small and acceptable. Of course for the success of the wave prediction the forcing from the atmospheric model is of crucial importance and we decided to verify also the wind speed forecast of ALADIN model comparing also with satellite measurements from ENVISAT and JASON1- see table 2 shows the results- notice the fact that there is an increase of the accuracy especially for wind speeds above 17m/s and overall the results confirms our experience that ARPEGE/ALADIN is a very reliable system for the Black Sea weather forecasts.

We also did a verification of SWAN, taking into account the satellite measurements of the significant wave height for locations with depth 10-20m, but selecting only points which are at least 10km far from the coast, because below 10km the coast can distort the measurements. Table 3 presents the results- the root mean square error is better than in deep waters, but the number of the measurements, on which the statistics are based is only 72, therefore it is still not possible to conclude that there is a better accuracy in the shelf area.

**Table 2.**

| ARPEGE/ ALADIN Wind speed forecast range | Bias [m/s] | Root Mean Square Error (RMSE) [m/s] | Scatter Index |
|---|---|---|---|
| **Day 1 (+6- +24h)** | -1.15 | 1.9 | 0.10 |
| **Day 2 (+24- +48h)** | -1.15 | 1.7 | 0.11 |
| **Day 3 (+48- +72h)** | -1.15 | 2.8 | 0.16 |
| **Entire period of forecast (+6- +72h)** | -1.15 | 2.3 | 0.13 |
| **Wind speed above 17m/s** | -1.1 | 1.7 | 0.07 |
| **Wind speed above 21m/s** | -1.0 | 1.1 | 0.02 |

**Table 3.**

| Significant Wave Height [m]- SWAN – shallow waters | Bias [m] | Root Mean Square Error (RMSE) [m] | Scatter Index |
|---|---|---|---|
| Entire period of forecast (+6- +72h) | -0.24 | 0.42 | 0.14 |

## DISCUSSION AND CONCLUSIONS

The overall conclusion is that SWAN model was found to produce a reliable wave forecasts for the Black Sea not only at deep, but also at coastal waters, but without operational measurements of the wave parameters from moored buoys the future progress will be difficult. Based on the conclusions from the stormy season 2011/2012 it was decided by NIMH- BAS to start a project to include the coastal warnings in the system METEOALARM (part of the European METEOALARM system), which provides warnings in case of dangerous meteorological and hydrological events.

## ACKNOWLEDGEMENTS

We thank Valery Spiridonov and Andrei Bogatchev for their efforts over the last decade to improve the Bulgarian ALADIN model and for the fruitful discussions. We also want to thank to Jean- Michel Lefevre from METEO- FRANCE for his advices. The satellite altimetry data and specific software tools was kindly provided by METEO FRANCE in the frame of bilateral cooperation between NIMH- BAS and METEO- FRANCE.